\documentclass[aps,preprint,showpacs,showkeys,nofootinbib,floatfix,superscriptaddress]{revtex4}
\usepackage{epsfig}
\usepackage{graphicx}
\usepackage{multirow}
\usepackage{amsmath,stackrel}
\usepackage{bm}
\newcommand{\be}{\begin{equation}}
\newcommand{\ee}{\end{equation}}
\newcommand{\beq}{\begin{equation}}
\newcommand{\eeq}{\end{equation}}
\newcommand{\bea}{\begin{eqnarray}}
\newcommand{\eea}{\end{eqnarray}}

\newcommand{\bk}{\bm k}

\newcommand{\ave}[1]{\langle {#1} \rangle}
\newcommand{\tave}[1]{\langle\!\langle{#1}\rangle\!\rangle}
\begin{document}
\title{Hadronic molecules with a ${\bar{D}}$ meson in a medium}

\author{T.~F.~Caram\'es}
\email{carames@usal.es}
\affiliation{Departamento de F\'\i sica Fundamental e IUFFyM, Universidad de Salamanca, E-37008
Salamanca, Spain}
\author{C.~E.~Fontoura}
\email{eduardo@ift.unesp.br}
\affiliation{Instituto Tecnol\'ogico de Aeron\'autica, DCTA, 12228-900 S\~ao Jos\'e dos Campos,
SP, Brazil}
\author{G.~Krein}
\email{gkrein@ift.unesp.br}
\affiliation{Instituto de F\'{\i}sica Te\'{o}rica, Universidade Estadual
Paulista, Rua Dr. Bento Teobaldo Ferraz, 271 - Bloco II, 01140-070 S\~ao Paulo, SP, Brazil}
\author{K.~Tsushima}
\email{kazuo.tsushima@gmail.com}
\affiliation{Laborat\'orio de F{\'\i}sica Te{\'o}rica e Computacional,
Universidade Cruzeiro do Sul, 01506-000, S\~{a}o Paulo, SP, Brazil}
\author{J.~Vijande}
\email{javier.vijande@uv.es}
\affiliation{Departamento de F\'{\i}sica At\'{o}mica, Molecular y Nuclear, Universidad de Valencia (UV)
and IFIC (UV-CSIC), E-46100 Valencia, Spain}
\author{A.~Valcarce}
\email{valcarce@usal.es}
\affiliation{Departamento de F\'\i sica Fundamental e IUFFyM, Universidad de Salamanca, E-37008
Salamanca, Spain}

\begin{abstract}
We study the effect of a hot and dense medium on the binding energy of hadronic molecules 
with open-charm mesons. We focus on a recent chiral quark-model-based prediction of a 
molecular state in the $N \bar D$ system. We analyze how the two-body thresholds and the 
hadron-hadron interactions are modified when quark and meson masses and quark-meson couplings 
change in a function of the temperature and baryon density according to predictions of the 
Nambu--Jona-Lasinio model. We find that in some cases the molecular binding is enhanced in 
medium as compared to their free-space binding. We discuss the consequences of our findings 
for the search for exotic hadrons in high-energy heavy-ion collisions as well as in the forthcoming 
facilities FAIR or J-PARC.
\end{abstract}

\pacs{14.40.Lb,12.39.Pn,12.40.-y,24.85.+p}

\keywords{Hadron molecules, Potential models, Medium effects, Chiral Symmetry} 
\maketitle

%
\section{Introduction}
\label{secI}

Recent developments in hadron physics have been motivated by the observation of 
exotic hadrons~\cite{Bramb,Esposito,exotic,Chen}. Most of them lie near open heavy-flavor thresholds 
implying that they may form the so-called hadronic molecules, that are colorless hadronic clusters
loosely bound by a relatively weak residual interaction. This could be a 
possible situation of what is expected to occur, in general, for multiquark systems. 
In this respect, the recently discovered five-quark baryonic resonances by the LHCb 
Collaboration at the Large Hadron Collider (LHC) at CERN, $P_c(4380)^+$ and 
$P_c(4450)^+$~\cite{Aai15}, have been interpreted as simple baryon-meson bound states. 
Also, some of the exotic mesonic states discovered in the hidden-charm and hidden-beauty 
sectors~\cite{Cho03,Aai14,Chi14,Bon12} might be well understood as meson-meson resonances. 
However, a comprehensive theoretical explanation of the nature of these exotic states is 
still missing~\cite{exotic,Bur15,Bra13}. 

The interest in structures containing hadrons with heavy flavors has also been 
recently reinvigorated by several studies about the existence of nuclear bound states
with heavy mesons~\cite{Dmesic-Tsu,Dmesic-Tol} and baryons~\cite{Tsu03,Tsu04,Gar15,Mae16}, 
the latter ones already predicted soon after the discovery of baryons possessing net 
charm~\cite{Dov77,Iwa77,Gat78}. In fact, there are theoretical estimations of the production 
cross sections as well as experimental requirements for producing charmed hypernuclei by 
means of charm exchange reactions on nuclei~\cite{Bre89}. Last but not least, it is also 
worth mentioning the suggestion of possible bound states of charmonium in nuclei due to multiple 
gluon exchange~\cite{Bro90,Bro97}. Such a possibility has been recently revisited by means of 
effective Lagrangians~\cite{Kre11,Tsu11} and effective Gaussian potentials~\cite{Yok12,Yok14}, 
stimulated by the lattice QCD suggestion of a weakly attractive interaction 
between charmonium and nucleons~\cite{Kaw10}.  

On the experimental side, there are exciting perspectives at extant and forthcoming facilities. 
At the LHC, all four collaborations, ALICE, ATLAS, CMS and LHCb, are engaged in searches for 
exotic hadrons. Particularly interesting is the possibility of the production of exotic hadrons 
in the hot and dense environment created in a high-energy heavy-ion collisions. In such an 
environment, heavy quarks are abundantly produced and they can pick up light-flavor quarks 
and antiquarks during the evolution of the medium and form multiquark states through a 
coalescence mechanism~\cite{SatoYazaki,Cho}. The production of bound states of $D$ mesons 
with nucleons and nuclei can be achieved in different laboratories worldwide. There are
planned experiments by the $\overline{\rm P}$ANDA Collaboration to produce them by annihilating
antiprotons on nuclei at the Facility for Antiproton Ion Research (FAIR)~\cite{Wie11,Hoh11}. 
The planned installation of a 50~GeV high-intensity proton beam at Japan Proton Accelerator 
Research Complex (J-PARC)~\cite{Tam12,Jparc,Tsu08} provides an additional opportunity. 
A Super$B$ collider~\cite{Fel12} offers similar possibilities. 

Thus, the physics of charm in the nuclear medium is becoming a hot topic and is expected to 
bring further progress in our understanding of the basic theory of the strong interaction, 
quantum chromodynamics (QCD). First-principles, analytical calculations within QCD of 
nuclear processes are presently impossible and, consequently, the interpretation of nuclear 
reaction results will always be afflicted by large uncertainties. The complete lack of 
experimental information on elementary interactions of charmed hadrons with nucleons in free 
space imposes additional difficulties in accessing in-medium effects. In order to make progress, 
one way to proceed, as advocated in Refs.~\cite{Hai07,Hai08,Hai09,Hai10,Car12}, is to use 
models constrained as much as possible by symmetry arguments, analogies with other similar 
processes, and the use of different degrees of freedom. This is particularly true when it comes 
to quark-model studies of the production of bound states in a hot and dense medium, as possible 
in-medium changes of the properties of the light constituent quarks must be taken into account. 

The behavior of quarkonia in a hot medium has attracted much interest. It was for instance
suggested a long time ago, that in a color-deconfined medium with a temperature above
a critical value, $T_c$, charmonia will melt due to the color Debye screening, and thus serve 
as a signal for the formation of quark-gluon plasma~\cite{Mat86}. For the bottomonium spectrum, a 
similar phenomenon may occur. Indeed, it was recently reported that bottomonium
spectra have received significant modifications when comparing their yield in proton-proton and 
Pb-Pb collisions at the LHC~\cite{Cha12}. The investigation of the nuclear medium effects on 
charmed hadron bound states is the main objective of this work. For this purpose we will select
a charmed hadron molecule whose existence is determined by pure quark effects, 
the $(I)J^P=(1)5/2^-$ $\Delta \bar D^*$ molecule~\cite{Car12}. The bound state is determined
by an attractive short-distance quark-exchange interaction, a feature due to the Pauli principle 
at the quark level that cannot be captured by an effective Lagrangian employing low-dimension 
hadronic operators. Given the prominent role played by quark-exchange effects in free space, 
we investigate the impact of in-medium changes in the parameters of the chiral quark model used 
in the evaluation of the binding energy of the molecule. To assess the required in-medium 
dependence on the constituent quark masses and the coupling constants of the light quarks to the 
$\pi$ and $\sigma$ mesons, we use the Nambu--Jona-Lasinio (NJL) model~\cite{Nam61,Nab61}. 

The paper is organized as follows. We use Sec.~\ref{secII} for describing
the in-medium dependence of the basic ingredients of the constituent chiral quark model:
quark and meson masses as well as quark-meson couplings within a 
NJL framework. In Sec.~\ref{secIII} we study the hadron masses in medium
by means of the modifications we have derived for the basic parameters of 
the quark model used. We discuss in Sec.~\ref{secIV} the in-medium hadron-hadron
interactions and we briefly revise the solution of the two-body bound-state 
problem looking for bound states. We present and discuss our results in Sec.~\ref{secV}.
Finally, in Sec.~\ref{secVI} we summarize our main conclusions.
 
%
%
\section{Medium dependence of quark and meson masses and quark-meson couplings}
\label{secII}

Within the perspective of the chiral quark model, changes in the masses of the light hadrons
and their mutual interactions at finite temperature ($T$) and baryon density ($\rho_B$) are 
driven by the change of the order parameter of dynamical chiral symmetry breaking, the quark
condensate. For sufficiently large values of $T$ and $\rho_B$, the (absolute value of the) 
in-medium light quark condensate, $\tave{\bar q q}$, becomes very small in the 
chiral limit, it can actually vanish. 
For zero baryon density, lattice QCD simulations~\cite{Baz11} at almost physical
pion masses ($m_\pi = 161$~MeV) have shown a drastic decrease of $|\tave{\bar q q}|$ around 
a temperature of $T_{\rm pc} = 154 \pm 9$~MeV. For finite baryon densities, the combined $T$ 
and $\rho_B$ behavior of $\tave{\bar q q}$ is presently unknown; the main reason for the lack 
of this knowledge is due to difficulties of using the Monte Carlo methods of lattice QCD due 
to the sign problem~\cite{Aar15}. On the other hand, for low $T$ and $\rho_B$, there are 
model-independent predictions~\cite{Ger89,Dru91,Coh92} for $\tave{\bar q q}$:
\begin{equation}
\frac{\tave{\bar q q}}{\ave{\bar q q}} = 1 -  \sum_{\rm h}
\frac{\Sigma_h}{f^2_\pi m^2_\pi} \,
\rho^{\rm h}_s
= 1 - \frac{T^2}{8f^2_\pi} - \frac{1}{3} \, \frac{\rho_B}{\rho_0} \, ,
\label{cond-Trho}
\end{equation}
where $\Sigma_h = m_q \, \partial m_h/\partial m_q$, $\rho^{\rm h}_s$ is the scalar 
density of hadron ${\rm h}$ in matter, $m_q$ is the current quark mass, 
$\langle\bar q q \rangle$ is the vacuum light quark condensate, $f_\pi$ is the pion leptonic 
decay constant, and $\rho_0$ is the baryon saturation density of nuclear matter. 

While changes in the constituent quark mass for low values of $T$ and $\rho_B$ 
could be directly related to the $T$ and $\rho_B$ dependence of the condensate, the
calculation of corresponding changes in the masses of the $\pi$ and $\sigma$ mesons
and their couplings to the constituent quarks requires a model. In this work we 
employ the NJL model; in addition to reproducing the result in Eq.~(\ref{cond-Trho}), 
its bosonized version with $\pi$ and $\sigma$ mesons~\cite{Egu76} has the same 
Yukawa quark-meson couplings as those in the chiral constituent quark model (CCQM) of 
Ref.~\cite{Car12}, and it gives very 
simple expressions for the masses and couplings (for reviews on this and other 
applications of the model in different problems in hadron and nuclear physics, see e.g. 
Refs.~\cite{Vog91,Kle92,Hat94,Bub05}). 

To make the paper self-contained and set the notation, we review the basic features of 
the NJL model relevant for our purposes here. The results we use are derived from the 
Lagrangian density: 
\begin{equation}
{\cal L}_{NJL} = \bar{q}\left(i\not\!\partial - m_q\right)q 
+  G \left[(\bar{q}q)^2 + (\bar{q} i\gamma_5 {\bm \tau}q)^2\right] \, ,
\end{equation}
where $m_q$ is the current quark mass; we will work in the isospin symmetric 
limit, $m_q = m_u = m_d$. At finite $T$ and $\rho_B$, the meson masses $m_\sigma$ and 
$m_\pi$ and the quark-meson coupling constants $g_{qq\sigma}$ and $g_{qq\pi}$, defined 
respectively as the poles of the meson propagators and their residues, are obtained from 
the  equations ($M=\sigma,\pi$)
\beq
1 - 2\,G\,\Pi_M(\omega^2=m^2_M) = 0, \hspace{1.0cm} 
g^2_{qqM} =\left[\frac{\partial\,\Pi_M(\omega^2)}{\partial\,\omega^2}
\right]^{-1}_{\omega^2 = m^2_M}\, ,
\label{mass-coup}     
\eeq
where $\Pi_M(\omega^2)$ is the meson polarization function 
\beq
\Pi_M(\omega^2) = 2N_c N_f \int \frac{d^3\bk}{(2\pi)^3}
\frac{1- \left[\,n^+_q(\bk)+ n^-_q(\bk)\,\right]}{E_{q}(\bk)}\,F_M(\omega^2) \, ,
\label{PiM}
\eeq
with $N_c=3$ and $N_f=2$ being the number of colors and flavors, $E_q(\bk)=(M^2_q+\bk^2)^{1/2}$ \, ,
\bea
F_\pi(\omega^2) = \frac{E^{2}_{q}(\bk)}{E^{2}_{q}(\bk) - \omega^2/4},
\hspace{0.6cm}
F_\sigma(\omega^2) = \frac{E^{2}_{q}(\bk) - M^2_q}{E^{2}_{q}(\bk) - \omega^2/4}\, ,
\label{Fpi_Fsigma}
\eea
and $M_q$ is the constituent quark mass, which is a solution of the gap equation that involves the
quark condensate $\tave{\bar q q}$:
\begin{eqnarray}
M_{q} &=& m_{q} + 4 G \tave{\bar q q} \, ,
\label{gap_equation} \\ 
\tave{\bar q q} &=& N_{c}N_{f} M_{q}\,
\int\,\frac{d^{3}\bk}{(2\pi)^{3}}\frac{1-\left[n^+_{q}(\bk) + n^-_{q}(\bk)\right]}
{E_{q}(\bk)} \, .
\label{qcond}
\end{eqnarray}
Here and in Eq.~(\ref{PiM}), $n^{\pm}_{q}(\bk)$ are the quark and antiquark Fermi-Dirac 
distributions
\beq
n^\pm_q(\bk)=\frac{1}{1+e^{\beta [E_q(\bk) \mp \mu_B]}} \, ,
\label{FD-dist}
\eeq
\begin{figure}[t]
\begin{center}
\vspace{-2.5cm}
\includegraphics[scale=0.8]{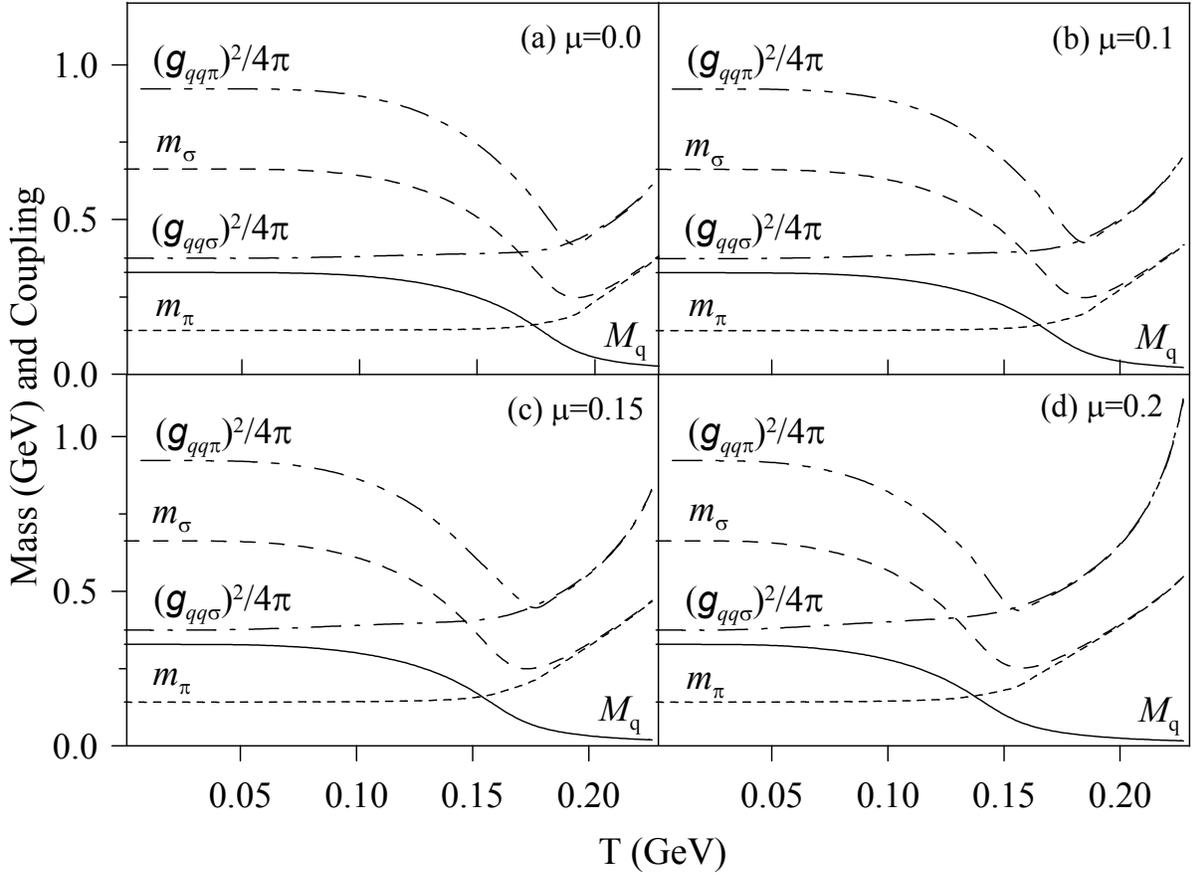}
\end{center}
\vspace{-9.cm}
\caption{Temperature dependence of the masses of the constituent quark $M_q$ (solid line), 
pion $m_\pi$ (dotted line), and the sigma $m_\sigma$ (dashed line), and the quark-meson 
couplings $g^2_{qq\sigma}/4\pi$ (dash-dotted line) and $g^2_{qq\pi}/4\pi$ (dash-double-dotted 
line) for different values of chemical potential $\mu$, in GeV.}
\label{fig1}
\end{figure}
with $\beta = k_B T$ and $\mu_B$ is the quark-baryon chemical potential. The baryon density 
$\rho_B$ is given in terms of these distributions by
\begin{equation}
\rho_{B} = \frac{2N_{c}N_{f}}{3}
\int\,\frac{d^{3}\bk}{(2\pi)^{3}}\left[n^+_q(\bk) - n^-_q(\bk)\right] \, .
\label{baryon_density}
\end{equation}

The integral in Eq.~(\ref{PiM}) is to be understood as a principal-value integral when 
$\omega^2 > 4 M^2_Q$. The temperature-independent part of the integrals in 
Eqs.~(\ref{PiM}) and (\ref{gap_equation}) are ultraviolet divergent and need regularization.
Since the model is nonrenormalizable, the regularization scheme is part of the model; here
we use a three-dimensional cutoff scheme parametrized by a cutoff~$\Lambda$. 

Vacuum quantities, which are used to fit the parameters of the model, are obtained from the 
above equations by setting the Fermi-Dirac distributions to zero. It is important to 
note that the couplings $g_{qq\sigma}$ and $g_{qq\pi}$ are not bare couplings; they incorporate 
the effects of dynamical chiral symmetry breaking (DCSB) and as such are different from each other. 
In the chiral quark model, such 
effects arise from corrections to the bare quark-meson vertices and are parametrized 
by phenomenological form factors. At high temperatures and densities, when chiral symmetry is
restored, the couplings $g_{qq\sigma}$ and $g_{qq\pi}$ become equal to each other, as we discuss 
in the next section. 

The free parameters of the  NJL model are the current quark mass $m_{q}$, the coupling 
$G$ and the cutoff~$\Lambda$. They are fixed by fitting the vacuum values for the quark 
condensate $\langle\bar{u}u\rangle=\langle\bar{d}d\rangle$, the pion decay constant $f_{\pi}$ and 
the pion mass $m_{\pi}$. Taking~\cite{Kle92} $m_{q} = m_{u} = m_{d} = 5$ MeV, $G\Lambda^2 = 2.14$ 
and $\Lambda = 653$ MeV, one obtains $\langle\bar{u}u\rangle=\langle\bar{d}d\rangle = 
- (252\,\,\text{MeV})^3$, $f_{\pi} = 94$~MeV and $m_{\pi} = 142$~MeV. With such parameters, 
one obtains for the constituent quark mass $M_{q} = M_{u} = M_{d} = 328$~MeV and 
for the $\sigma$ mass $m_{\sigma} = 663$~MeV. 

In Fig.~\ref{fig1} we present the results for the masses of the constituent quarks $M_{q}$, 
the $\pi$ and $\sigma$ masses, $m_M, M=\pi,\sigma$, and quark-meson couplings $g_{Mqq}$ 
as a function of temperature for different values of the chemical potential $\mu$, 
i.e., (a)\,$\mu=0$, (b)\,$\mu=0.1$\,GeV, (c)\,$\mu=0.15$\,GeV, and 
(d)\,$\mu=0.2$\,GeV. As expected, the constituent quark mass $M_q$ and the $\sigma$ 
mass $m_\sigma~(\sim~2 M_q)$ drop significantly at sufficiently high temperatures
and densities, while the couplings become degenerate. We also note that the vacuum values
of the masses of the constituent quarks and of the mesons differ from those in the CCQM
{\textemdash} see Table~II of Ref.~\cite{Car12}{\textemdash}by less 
than~5\%. One could readjust the parameters of the NJL model to obtain even closer results but, as
we are mainly interested in the medium dependence of these quantities, we simply use 
the ratios of the medium to vacuum values of the masses and couplings 
for calculating the baryon and meson masses and their interactions. 

We close this section by reflecting on the limitations and advantages of the present calculation. 
Initially, it should be clear that both the NJL and chiral quark models are supposed to
describe the chiral aspects of QCD in vacuum and at low temperature and baryon chemical potential. 
At sufficiently high temperature and chemical potential, chiral symmetry is restored and there is no quark condensate, constituent quarks, bound sigma mesons, effective quark-meson couplings,
etc. The models break down and are not adequate to describe QCD in such regimes. The discrete 
poles in the $\sigma$ and $\pi$ meson 
correlation functions melt into a continuum above the (pseudo) critical temperature, describing 
correlations of essentially massless quark-antiquark pairs with thermal masses that grow with
temperature. The NJL model is not expected to describe the precise QCD behavior of these thermal 
masses and it would be desirable to employ a model that interpolates between the low- and high-energy 
regimes of QCD, like those that incorporate confinement, dynamical chiral symmetry breaking and 
asymptotic freedom, and describe baryon bound states~\cite{Kre-Sea1,Kre-Sea2}. Such models would 
allow one to calculate all meson-baryon properties, in vacuum and in medium, within a single framework.
Medium effects can be incorporated in the spirit of the quark-meson coupling model, in which 
meson mean fields couple directly to current quarks in the hadron; for a review, 
see Ref.~\cite{QMC}. However, the study of hadron-hadron interactions, taking into account, 
in particular, quark-exchange effects, is difficult due to the use of an underlying soliton or 
bag model. In the context of a CCQM, an earlier investigation on nucleon and nuclear matter
properties, including quark exchange effects, is the one of Ref.~\cite{QMC-CQM}, but no application 
of the model to in-medium charmed hadrons is available. In view of this, and given the close 
relationship between the CCQM and the NJL model, we believe that our calculation captures 
the basic physics of chiral symmetry on quark exchange in the interaction of $\bar D$ mesons 
with nucleons.

%
%
\section{Hadron masses in medium}
\label{secIII}

In the CCQM of Ref.~\cite{Car12}, baryons are described as clusters of three interacting 
massive (constituent) quarks, with their mass coming from dynamical chiral symmetry breaking 
in QCD. Short-distance perturbative QCD effects are taken into account through the 
one-gluon exchange (OGE). In addition to the masses for the constituent quarks, DCSB implies 
the presence of (pseudo-) Goldstone bosons; their effects are taken into account by introducing 
them as explicit degrees of freedom via $\pi$ and $\sigma$ fields. These fields introduce 
long-range interactions between the light $u$ and $d$ constituent quarks. Quark confinement 
is incorporated via an effective potential that contains string-breaking effects. The charm and 
light quarks interact only via one-gluon exchange and, of course, are subject to the same
confining potential. For a review on the model as well as the technical details and methods to 
solve the two- and three-quark problems, see Refs.~\cite{Val05,Vij05}.

\begin{figure*}[t]
\resizebox{8.15cm}{13.cm}{\includegraphics{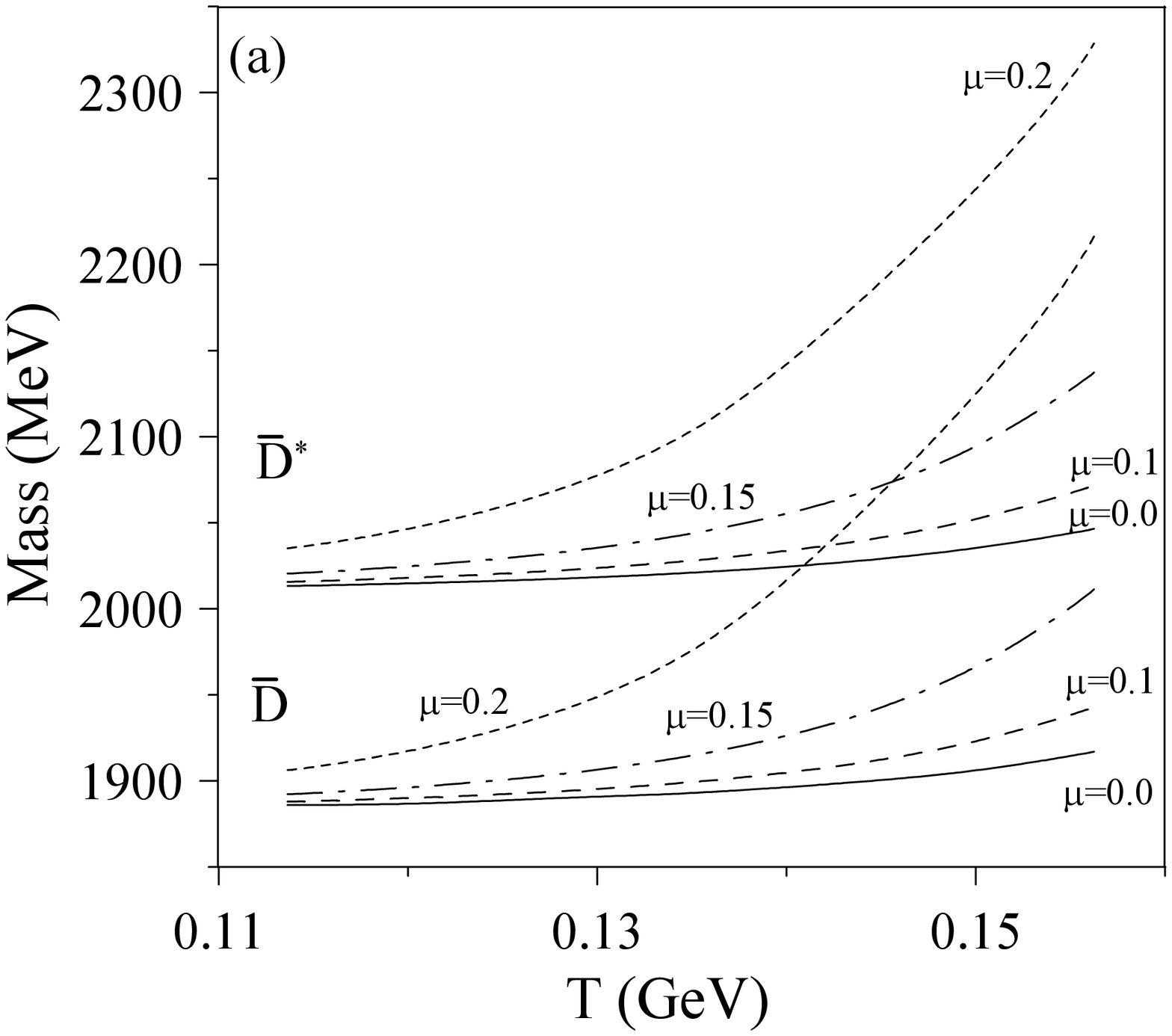}}
\resizebox{8.15cm}{13.cm}{\includegraphics{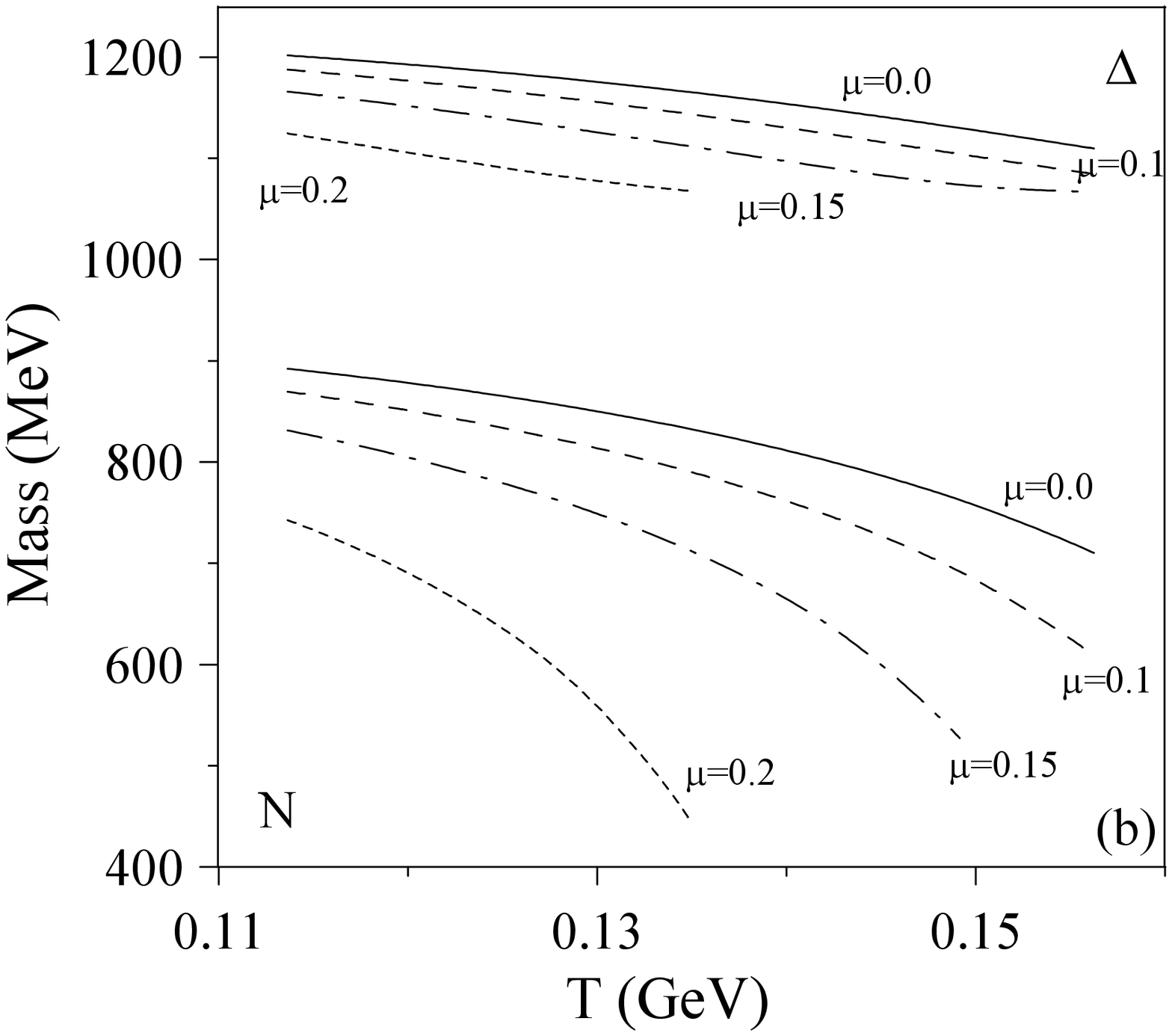}}
\vspace*{-6.5cm}
\caption{(a) Masses of the $\bar D$ and $\bar D^*$ mesons as a function of 
the temperature for the different chemical potentials. (b) Same as (a) for the
$N$ and $\Delta$ baryons.}
\label{fig2}
\end{figure*}

Given the temperature and chemical potential dependence of quark and meson masses and quark-meson
couplings derived in the previous section, one can calculate the masses of the hadrons 
of interest: $\bar D$ and $\bar D^*$ mesons and $N$ and $\Delta$ baryons. 
We show in Fig.~\ref{fig2}(a) the variation of the $\bar D$ and $\bar D^*$ meson masses as a 
function of the temperature for the different chemical potentials. There is almost no variation
of the mass for any chemical potential for a temperature below 0.1 GeV. For temperatures above
this value, the masses of the pseudoscalar and vector mesons change in a rather similar manner, 
which makes manifest that the variation of the mass is a spin-independent effect. 
Being systems made of a light and a heavy quark, only confinement and the one-gluon exchange
contribute to the mass of the $\bar D$ and $\bar D^* $ mesons. The dominant contribution to the 
variation of the masses with temperature comes from the kinetic energy due to the reduction of the 
mass of the light constituent quark. The mass of the charm quark is not modified by temperature, and  
the reduced mass of the heavy-light system approaches the mass of the light quark.
It is therefore the kinetic energy that is mainly responsible for the change in the mass of the mesons. 
The modifications in the spin-dependent part, which are responsible for the mass difference between the
$\bar D$ and the $\bar D^*$ mesons, are minimized due to the presence of the heavy-quark mass in the 
denominator of the one-gluon exchange through its $1/\left( m_{q_i} \, m_{q_j}\right)$ dependence.
Similar results have been obtained in the literature very recently for the variation of the 
$\bar D$ meson mass in the nuclear medium using QCD sum rules~\cite{Suz16}.

In Fig.~\ref{fig2}(b) we depict the variation of the masses of the $N$ and $\Delta$ baryons as 
a function of the temperature for the different chemical potentials. As we can see there are 
important differences as compared to the $\bar D$ meson case, due to the presence of three 
light quarks. Being a more involved system, it can be easily concluded that the diminishing of 
the mass of the nucleon is mainly due to the spin-dependent part of the one-gluon exchange, which 
also generates the decreasing of the mass of the $\Delta$. While the 
$(\vec \sigma \cdot \vec \sigma)(\vec \tau \cdot \vec \tau)$ structure of the 
chiral pseudoscalar interaction gives attraction for symmetric spin-isospin pairs and repulsion for 
antisymmetric ones, which would augment the mass of the $\Delta$, the 
$(\vec \sigma \cdot \vec \sigma)(\vec \lambda \cdot \vec \lambda)$ structure
of the color-magnetic part of the OGE gives similar contributions in both cases, diminishing
the mass of the $N$ and the $\Delta$. The effect is much more pronounced in the case of
the $N$, due to the presence of a spin-zero diquark, where the OGE is attractive and 
this effect is increased when the mass of the quark is diminished, as it happens when the
temperature and the chemical potential change. 

We note that in principle the OGE and confining potentials in the chiral quark model
are temperature dependent due to Debye screening, as demonstrated e.g. by a recent lattice
QCD calculation in Ref.~\cite{Latt-Deb}. However, such a temperature dependence has a minor 
impact on our calculations. This is because strong modifications of the potentials appear at 
long distances and only for temperatures well above the critical temperature, as shown in
Fig.~10 of Ref.~\cite{Latt-Deb}. In fact, this has been analyzed in a phenomenological manner
by some of the present authors in Ref.~\cite{Screen-us}, where it was shown that the quarkonia
ground-state masses are almost independent of the temperature until very close to the critical 
temperature, above which the hadrons melt. For even larger temperatures, as discussed in the 
previous section, the underlying models we use lose applicability. Moreover, as explained in 
detail in Ref.~\cite{Car12}, in order to evaluate the interaction kernel between two hadrons
(see next section) one must subtract the self-energy contributions 
from the kernel and, consequently, any possible modification of confinement would drop out in 
the calculation of the interaction between the two hadrons.

Once we have determined the effect of the temperature and the chemical potential
on the hadron masses, we know the thresholds for the study of the possible
existence of meson-baryon resonances in nuclear matter.

%
%
\section{In-medium binding of $N\bar D$ molecules}
\label{secIV}
\begin{figure}[b]
\vspace*{-2.5cm}
\hspace*{-1cm}\mbox{\epsfxsize=100mm\epsffile{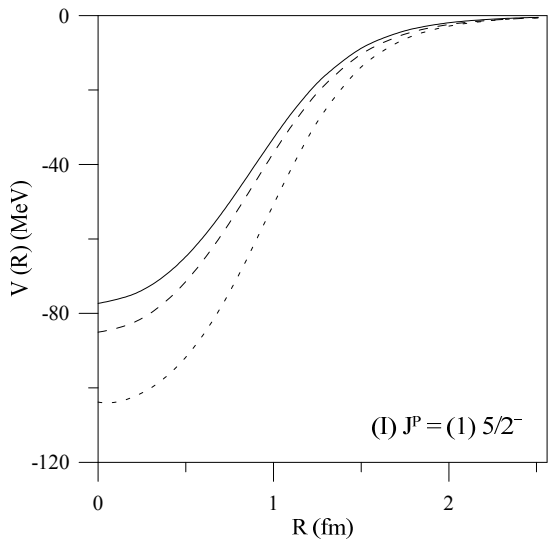}}
\vspace*{-3.5cm}
\caption{$(I,J)=(1,5/2)$ $\Delta \bar D^*$ interaction 
as a function of the temperature for different values of the chemical potential.
The solid line stands for the free case $(T,\mu)=(0.,0.)$, the dashed line for
$(T,\mu)=(0.12,0.)$, and the dotted line for $(T,\mu)=(0.12,0.15)$, where $T$ and
$\mu$ are given in GeV.}
\label{fig3}
\end{figure}
Next, we investigate how the hadron-hadron interactions are modified
in a medium at finite $T$ and $\mu$. For this purpose we follow exactly the same scheme 
that has been detailed in Ref.~\cite{Car12}, evaluating the interacting potentials with the 
quark and meson masses and quark-meson coupling constants determined in Sec.~\ref{secII} 
for the different temperatures and chemical potentials. We note that our calculation
is particularly applicable for a medium similar to the one formed in a high-energy 
heavy-ion collision, in which quarks coalesce to form weakly bound hadron molecules~\cite{Cho}. 
We will center our attention on the particular state highlighted in Ref.~\cite{Car12}, the 
$\Delta \bar D^*$ state with 
$(I)J^P=(1)5/2^-$. We show in Fig.~\ref{fig3} the $\Delta \bar D^*$ interaction with 
$(I,J)=(1,5/2)$ for some selected values of the temperature and the chemical potential. 
As we can see, the effect of the medium is to strengthen the interaction. This is due
to the fact that the interaction in this channel is controlled by the
scalar exchange due to the almost exact cancellation of the repulsive one-gluon
exchange and the attractive one-pion exchange [see Fig. 3(b) of Ref.~\cite{Car12}].
When increasing either the temperature or the chemical potential, the interactions
grow, but the cancellation between the repulsive one-gluon exchange and the 
attractive one-pion exchange still remains, and the diminishing of the mass 
of the scalar boson with the temperature and the chemical potential, generates 
a stronger interaction.

To study the possible existence of an exotic state in this particular channel
in the medium, we solve the Lippmann-Schwinger equation for negative 
energies by looking at the Fredholm determinant $D_F(E)$ at zero energy~\cite{Gar87}. 
If there are no interactions then $D_F(0)=1$, 
if the system is attractive then $D_F(0)<1$, and if a
bound state exists then $D_F(0)<0$. 
We consider a baryon-meson system $Q_i R_j$ ($Q_i=N$ or $\Delta$ and $R_j=\bar D$ 
or $\bar D^*$) in a relative $S$ state 
interacting through a potential $V$ that contains a
tensor force. Then, in general, there is a coupling to the 
$Q_i R_j$ $D$ wave. Moreover, the baryon-meson system could couple to other 
baryon-meson states, $Q_k R_m$ ( in the present case there would not
be coupling between different physical systems). If we denote the different 
baryon-meson systems as channel $A_i$, the Lippmann-Schwinger equation for the baryon-meson 
scattering becomes
\begin{eqnarray}
t_{\alpha\beta;IJ}^{\ell_\alpha s_\alpha, \ell_\beta s_\beta}(p_\alpha,p_\beta;E)& = & 
V_{\alpha\beta;IJ}^{\ell_\alpha s_\alpha, \ell_\beta s_\beta}(p_\alpha,p_\beta)+
\sum_{\gamma=A_1,A_2,\cdots}\sum_{\ell_\gamma=0,2} 
\int_0^\infty p_\gamma^2 dp_\gamma V_{\alpha\gamma;IJ}^{\ell_\alpha s_\alpha, \ell_\gamma s_\gamma}
(p_\alpha,p_\gamma) \nonumber \\
& \times& \, G_\gamma(E;p_\gamma)
t_{\gamma\beta;IJ}^{\ell_\gamma s_\gamma, \ell_\beta s_\beta}
(p_\gamma,p_\beta;E) \,\,\,\, , \, \alpha,\beta=A_1,A_2,\cdots \,\, ,
\label{eq0}
\end{eqnarray}
where $t$ is the two-body scattering amplitude, $I$, $J$, and $E$ are the
isospin, total angular momentum and energy of the system,
$\ell_{\alpha} s_{\alpha}$, $\ell_{\gamma} s_{\gamma}$, and
$\ell_{\beta} s_{\beta }$
are the initial, intermediate, and final orbital angular momentum and spin, respectively,
 and $p_\gamma$ is the relative momentum of the
two-body system $\gamma$. The propagators $G_\gamma(E;p_\gamma)$ are given by
\begin{equation}
G_\gamma(E;p_\gamma)=\frac{2 \mu_\gamma}{k^2_\gamma-p^2_\gamma + i \epsilon} \, ,
\end{equation}
with
\begin{equation}
E=\frac{k^2_\gamma}{2 \mu_\gamma} \, ,
\end{equation}
where $\mu_\gamma$ is the reduced mass of the two-body system $\gamma$.
For bound-state problems $E < 0$ so that the singularity of the propagator
is never touched and we can forget the $i\epsilon (\epsilon > 0)$ in the denominator.
If we make the change of variables
\begin{equation}
p_\gamma = d\, \frac{1+x_\gamma}{1-x_\gamma},
\label{eq2}
\end{equation}
where $d$ is a scale parameter, and the same for $p_\alpha$ and $p_\beta$, we can
write Eq.~(\ref{eq0}) as
\begin{eqnarray}
t_{\alpha\beta;IJ}^{\ell_\alpha s_\alpha, \ell_\beta s_\beta}(x_\alpha,x_\beta;E)& = & 
V_{\alpha\beta;IJ}^{\ell_\alpha s_\alpha, \ell_\beta s_\beta}(x_\alpha,x_\beta)+
\sum_{\gamma=A_1,A_2,\cdots}\sum_{\ell_\gamma=0,2} 
\int_{-1}^1 d^2\left(\frac{1+x_\gamma}{1-x_\gamma} \right)^2 \,\, \frac{2d}{ (1-x_\gamma)^2}\,
dx_\gamma \nonumber \\
&\times & V_{\alpha\gamma;IJ}^{\ell_\alpha s_\alpha, \ell_\gamma s_\gamma}
(x_\alpha,x_\gamma) \, G_\gamma(E;p_\gamma) \,
t_{\gamma\beta;IJ}^{\ell_\gamma s_\gamma, \ell_\beta s_\beta}
(x_\gamma,x_\beta;E) \, .
\label{eq3}
\end{eqnarray}
We solve this equation by replacing the integral from $-1$ to $1$ by a
Gauss-Legendre quadrature which results in the set of
linear equations
\begin{equation}
\sum_{\gamma=A_1,A_2,\cdots}\sum_{\ell_\gamma=0,2}\sum_{m=1}^N
M_{\alpha\gamma;IJ}^{n \ell_\alpha s_\alpha, m \ell_\gamma s_\gamma}(E) \, 
t_{\gamma\beta;IJ}^{\ell_\gamma s_\gamma, \ell_\beta s_\beta}(x_m,x_k;E) =  
V_{\alpha\beta;IJ}^{\ell_\alpha s_\alpha, \ell_\beta s_\beta}(x_n,x_k) \, ,
\label{eq4}
\end{equation}
with
\begin{eqnarray}
M_{\alpha\gamma;IJ}^{n \ell_\alpha s_\alpha, m \ell_\gamma s_\gamma}(E)
& = & \delta_{nm}\delta_{\ell_\alpha \ell_\gamma} \delta_{s_\alpha s_\gamma}
- w_m d^2\left(\frac{1+x_m}{1-x_m}\right)^2 \frac{2d}{(1-x_m)^2} \nonumber \\
& \times & V_{\alpha\gamma;IJ}^{\ell_\alpha s_\alpha, \ell_\gamma s_\gamma}(x_n,x_m) 
\, G_\gamma(E;{p_\gamma}_m),
\label{eq5}
\end{eqnarray}
and where $w_m$ and $x_m$ are the weights and abscissas of the Gauss-Legendre
quadrature while ${p_\gamma}_m$ is obtained by putting
$x_\gamma=x_m$ in Eq.~(\ref{eq2}).
If a bound state exists at an energy $E_B$, the determinant of the matrix
$M_{\alpha\gamma;IJ}^{n \ell_\alpha s_\alpha, m \ell_\gamma s_\gamma}(E_B)$ 
vanishes, i.e., $\left|M_{\alpha\gamma;IJ}(E_B)\right|=0$.
We took the scale parameter $d$ of Eq.~(\ref{eq2}) as $d=$ 3 fm$^{-1}$
and used a Gauss-Legendre quadrature with $N=$ 20 points.
\begin{figure}[t]
\vspace*{-4cm}
\mbox{\epsfxsize=150mm\epsfysize=190mm\epsffile{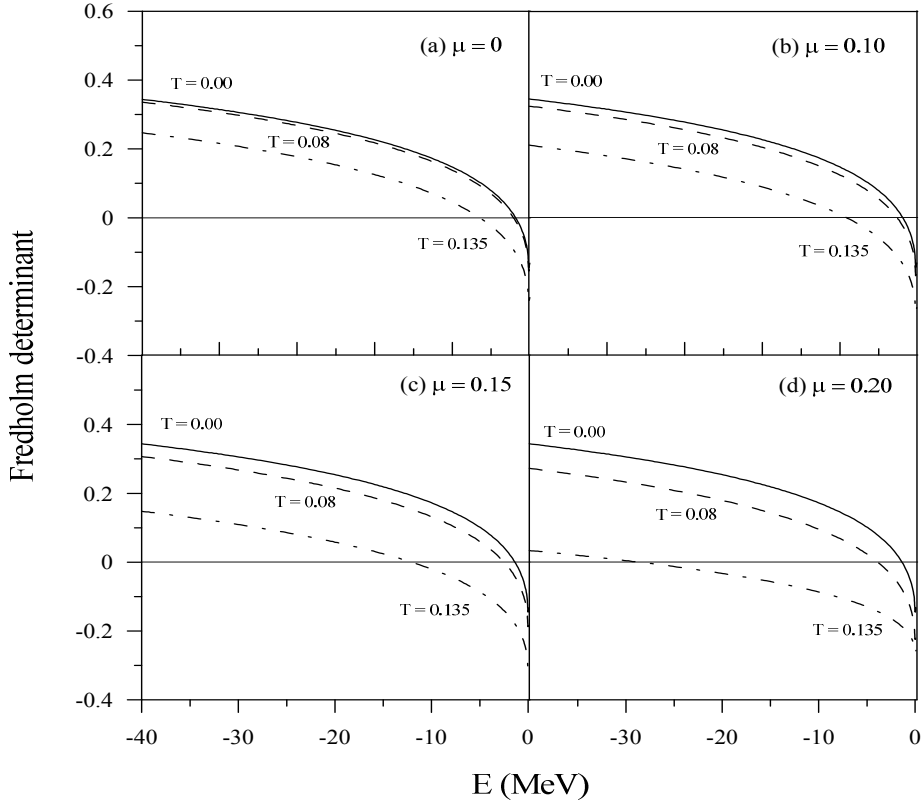}}
\vspace*{-4.5cm}
\caption{$(I)J^P=(1)5/2^-$ $\Delta \bar D^*$ Fredholm determinant for selected
temperatures of the different chemical potentials (in GeV).}
\label{fig4}
\end{figure}
%
%
\section{Results and discussion}
\label{secV}
\begin{figure}[t]
\vspace*{-2.5cm}
\hspace*{-1cm}\mbox{\epsfxsize=100mm\epsffile{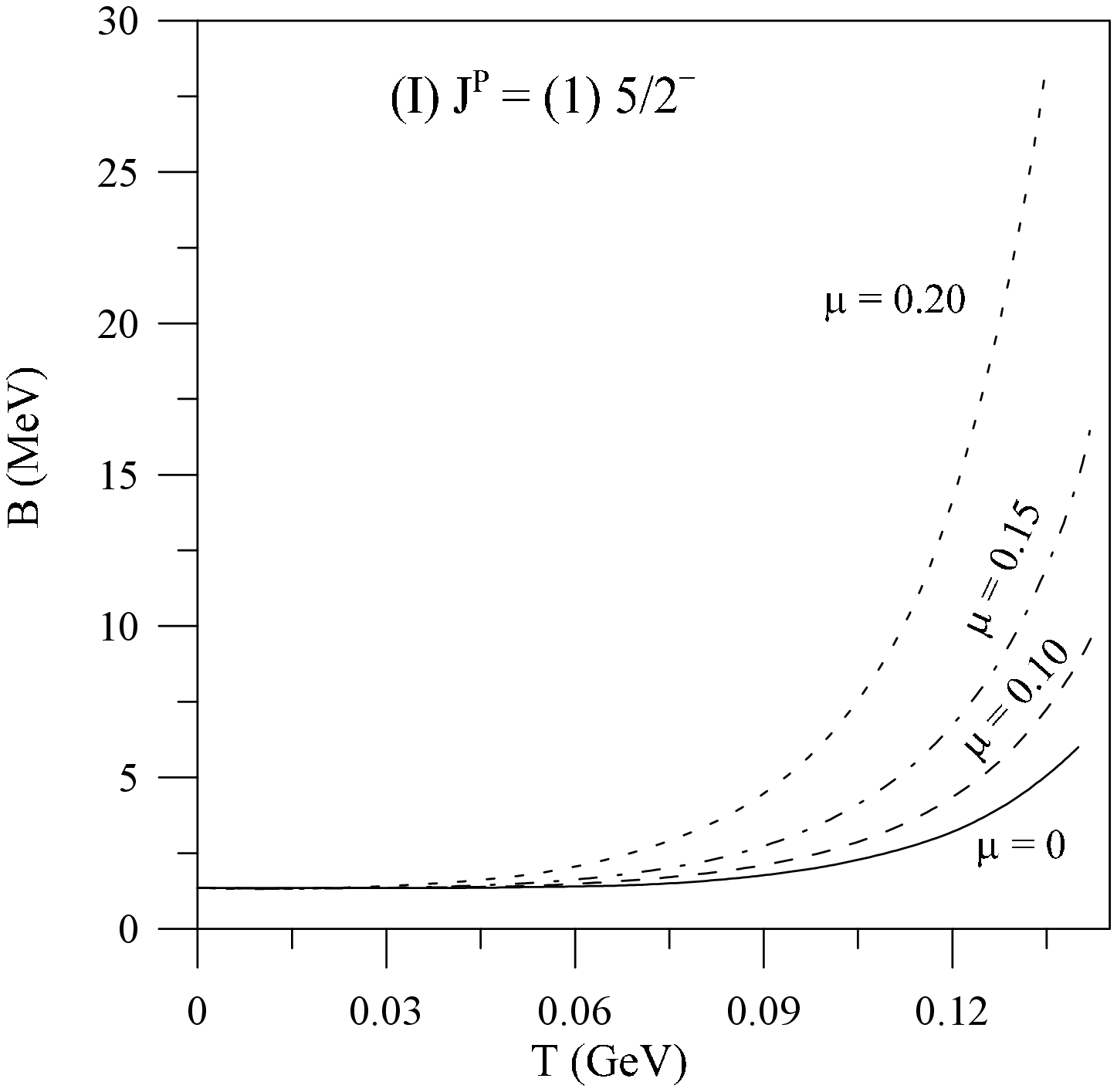}}
\vspace*{-3.5cm}
\caption{Binding energy of the $(I)J^P=(1)5/2^-$ $\Delta \bar D^*$ bound state, 
as a function of the temperature for different values of the chemical potential (in GeV).}
\label{fig5}
\end{figure}

The existence of charmed hadron molecules has been a topic of interest
in recent years in different theoretical frameworks, as chiral 
quark-models~\cite{Car12,Fon13}, boson-exchange models~\cite{Dot13},
or effective Lagrangian approaches~\cite{Yas09,Yam11}. As already mentioned, our interest
here is in the charmed hadron molecule $\Delta\bar{D}^*$ with isospin-spin 
quantum numbers $(I,J)=(1,5/2)$ that was recently predicted~\cite{Car12} within a chiral 
constituent quark model approach~\cite{Val05,Vij05}. Our interest is motivated 
mainly by the crucial role played by an attractive short-distance quark-exchange 
interaction, which is a prominent feature due to the Pauli principle at the quark level. This is
important because, in general, quark-exchange effects cannot be captured by an effective 
Lagrangian employing low-dimension hadronic operators. This feature was explicitly demonstrated in 
Ref.~\cite{Car12} for the case of the $N\bar D$ system by comparing predictions from the chiral 
quark model~\cite{Val05,Vij05}, and an effective Lagrangian~\cite{Yas09,Yam11} satisfying 
heavy-quark and chiral symmetries. Given the prominent role played by quark-exchange effects in 
free space, and the possibility that such a molecule can be formed in the environment of 
a heavy-ion collision~\cite{Cho}, it is important to investigate the impact of in-medium 
changes in the parameters of the chiral quark model used in the evaluation of the binding 
energy of the molecule. The implications of our results in the coalescence dynamics in
the formation of the molecule is left for a future publication. 

Making use of the in-medium hadron masses and hadron-hadron interactions 
derived in the previous section, we have solved the Lippmann-Schwinger 
equation for the $(I)J^P=(1)5/2^-$ $\Delta \bar D^*$ system. 
We show in Fig.~\ref{fig4} the Fredholm determinant for selected
temperatures of the different chemical potentials. In all cases $E=0$
corresponds to the mass of the corresponding threshold, i.e. 
$M_{\bar D^*}(T,\mu)+M_{\Delta}(T,\mu)$. As pointed out in the determination
of the in-medium masses of $\bar D$ mesons and $N$ and $\Delta$'s,
there is almost no variation of the binding energy
for any chemical potential for a temperature below 0.1 GeV.
For temperatures above this value it can be seen that the 
binding energy increases when increasing, the temperature and/or the chemical
potential. This is so in spite of the fact that the mass of the threshold 
diminishes, increasing in this way the kinetic energy. However the change
of the interacting potential due to the diminishing of the mass of the
scalar boson is capable of increasing the binding. In Fig.~\ref{fig5} we show
the binding energy of the $(I)J^P=(1)5/2^-$ $\Delta \bar D^*$ state, 
as a function of the temperature for different values of the chemical potential.
It varies between 1.35 MeV for the free case up to around 30 MeV for the
harder system we have considered.

The bound state found in the $(I,J)=(1,5/2)$ $\Delta \bar D^*$ channel would 
also appear in the scattering of $D$ mesons on nucleons as a D-wave resonance, which 
could in principle be measured in the near future. There are proposals for
experiments by the $\overline{\text{P}}$ANDA Collaboration~\cite{Wie11,Hai08} to produce $D$ mesons 
by annihilating antiprotons on the deuteron. They are based on recent estimations of 
the cross section for the production of $D \bar D$ pairs in proton-antiproton 
collisions~\cite{Kho12,Hai14}. The predicted bound state has quantum numbers
$(I)J^P=(1)5/2^-$ and it constitutes a sharp prediction of quark-exchange 
dynamics because in a hadronic model the attraction appears in different 
channels~\cite{Yas09,Yam11}.

The charmed molecule discussed in this paper, reflects the possible importance of quark
antisymmetrization dynamics, which was already noted when studying the $\Delta\Delta$ system within 
the CCQM model~\cite{Val01}. In that case, an S-wave resonance with maximum spin was 
predicted. Experimental evidence of such a resonance was found in the $NN$ scattering data, 
in the $^3D_3$ partial wave, and therefore as a D-wave resonance in the $NN$ scattering~\cite{Arn00}. 
This finding, the so-called $d^*(2380)$ resonance, has been used by the CELSIUS/WASA 
Collaboration~\cite{Bas09} as a possible explanation of the cross section of the double-pionic
fusion reaction $pn \rightarrow d\pi^0 \pi^0$. The interpretation of the CELSIUS/WASA data
is compatible with the formation of a $\Delta\Delta$ intermediate state, which also provides
a plausible explanation for the Abashian, Booth and Crowe effect~\cite{Aba60}:
an unexpected enhancement in the double-pionic fusion of deuterons and protons to $^3{\rm He}$.

%
%
\section{Summary}
\label{secVI}

In brief, we have studied the effects of temperature and baryon density on the binding energy
of hadronic molecules with open-charm mesons. We have selected a recent chiral 
quark-model-based prediction of a resonance in the $\Delta \bar D^*$ system with quantum 
numbers $(I)J^P=(1)5/2^-$. We have analyzed the modification of the thresholds 
and the hadron-hadron interactions when the basic parameters of the CCQM, which are the quark 
and meson masses and quark-meson couplings, change with temperature and baryon density as 
predicted by the Nambu--Jona-Lasinio model. We have found that the in-medium binding energy 
is enhanced as compared to its free-space binding by the presence of the hot and dense matter. 
This finding is relevant for ongoing heavy-ion experiments at the Relativistic Heavy-Ion Collider
(RHIC) and the LHC, and for the planned experiments at FAIR and J-PARC.

As already mentioned, exotic hadrons like the hadron molecule $\Delta \bar D^*$ we studied 
here can be produced and realistically measured in high-energy heavy-ion collisions. 
In such collisions a hot and dense medium is created with abundant production of heavy 
quarks that, during the evolution of the medium, can pick up light-flavor quarks and 
antiquarks and form multiquark states through a coalescence mechanism~\cite{SatoYazaki,Cho}. 
Quark coalescence occurs during the phase transition to the hadronic phase and it has been
proven to give a very successful description of particle production in heavy-ion collisions
at RHIC~\cite{Fries}. Coalescence depends very sensitively on the spatial extension 
of the produced hadrons~\cite{SatoYazaki} and as such, modifications to the binding energy
of hadrons, in particular of molecular states, due to temperature and density will obviously
affect their production rates in heavy-ion collisions. The implications of our results in the 
coalescence dynamics in the formation of the $\Delta \bar D^*$ molecule is out of the scope 
of the present paper and is left for a future publication. 

The future research programs at different facilities like FAIR and J-PARC are expected to 
shed light on the uncertainties of our knowledge about the in-medium dependence of the 
hadron-hadron interaction with heavy flavors. While the scarce experimental information
leaves room for some degree of speculation in the study of processes involving charmed hadrons, 
the situation can be ameliorated with the use of 
well-constrained models based as much as possible 
on symmetry principles and analogies with other similar processes. The present detailed theoretical 
investigation of the possible existence of bound states in a hot and dense medium is based
on well-established models which provide the basic tools to make progress in the knowledge of 
properties of the $N \bar D$ interaction. It is hoped that our work is of help toward raising 
the awareness of experimentalists that it is worthwhile to investigate few-baryon systems, specifically 
because for some quantum numbers such states could be stable. 

%
%
\section*{Acknowledgments}
This work has been partially funded  by Ministerio de Educaci\'on y Ciencia and EU FEDER under 
Contracts No. FPA2013-47443 and FPA2015-69714-REDT, by Junta de Castilla y Le\'on under
Contract No. SA041U16, and by a bilateral agreement Universidad de 
Salamanca - Funda\c{c}\~ao de Amparo \`a Pesquisa do Estado de S\~ao Paulo - FAPESP Grant 
No. 2015/50326-5.  T.F.C. and A.V. are thankful for financial support from the Programa Propio 
XIII of the University of Salamanca. Partial financial support is also acknowledged from Conselho 
Nacional de Desenvolvimento Cient\'{\i}fico e Tecnol\'ogico - CNPq, Grants No. 150659/2015-6 (C.E.F.), 
305894/2009-9 (G.K.), 400826/2014-3 and 308088/2015-8 (K.T.), and Funda\c{c}\~ao de Amparo \`a 
Pesquisa do Estado de S\~ao Paulo - FAPESP, Grants No. 2013/01907-0 (G.K.) and 2015/17234-0 (K.T.).C. E. F. is thankful for financial support from the Coordena\c{c}\~ao  de Aperfei\c{c}oamento de Pessoal de N\'ivel Superior - Capes, Grant No. 0251/14-3.

%
%


\begin{thebibliography}{99}

\bibitem{Bramb} N.~Brambilla {\it et al.},  
                 Eur. Phys. J. C {\bf 74}, 2981 (2014).

\bibitem{Esposito} A.~L.~Guerrieri, F.~Piccinini, A.~Pilloni, and A.~D.~Polosa,
                 Int. J. Mod. Phys. A {\bf 30}, 1530002 (2015).       

\bibitem{exotic} R.~A.~Brice\~no {\it et al.},
                 Chin. Phys. C {\bf 40}, 042001 (2016).
       
\bibitem{Chen} H.~X.~Chen, E.~L.~Cui, W.~Chen, X.~Liu, and S.~L.~Zhu,
                 arXiv:1606.03179.                        

\bibitem{Aai15} R.~Aaij {\it et al.} (LHCb Collaboration),
								Phys. Rev. Lett. {\bf 115}, 072001 (2015).
								
\bibitem{Cho03} S.~K.~Choi {\it et al.} (Belle Collaboration), 
								Phys. Rev. Lett. {\bf 91}, 262001 (2003).

\bibitem{Aai14} R.~Aaij {\it et al.} (LHCb Collaboration),
								Phys. Rev. Lett. {\bf 112}, 222002 (2014).

\bibitem{Chi14} K.~Chilikin {\it et al.} (Belle Collaboration),
								Phys. Rev. D {\bf 90}, 112009 (2014).

\bibitem{Bon12} A.~Bondar {\it et al.} (Belle Collaboration),
								Phys. Rev. Lett. {\bf 108}, 122001 (2012).

\bibitem{Bur15} T.~J.~Burns,
								Eur. Phys. J. A {\bf 51}, 152 (2015).

\bibitem{Bra13} E.~Braaten,
								Phys. Rev. Lett. {\bf 111}, 162003 (2013).

\bibitem{Dmesic-Tsu} K.~Tsushima, D.~H.~Lu, A.~W.~Thomas, K.~Saito, and R.~H.~Landau,
                                Phys. Rev. C {\bf 59}, 2824 (1999).

\bibitem{Dmesic-Tol} L.~Tolos,
                                Int. J. Mod. Phys. E {\bf 22}, 1330027 (2013).

\bibitem{Tsu03} K.~Tsushima and F.~C. Khanna,
								Phys. Rev. C {\bf 67}, 015211 (2003).
								
\bibitem{Tsu04} K.~Tsushima and F.~C. Khanna,
								J. Phys. G {\bf 30}, 1765 (2004).
								
\bibitem{Gar15} H.~Garcilazo, A.~Valcarce, and T.~F.~ Caram\'es,
								Phys. Rev. C {\bf 92}, 024006 (2015).

\bibitem{Mae16} S.~Maeda, M.~Oka, A.~Yokota, E.~Hiyama, and Y.-R.~Liu,
								Prog. Theor. Exp. Phys. {\bf 2016}, 023D02 (2016). 
												
\bibitem{Dov77} C.~B.~Dover and S.~H.~Kahana, 
								Phys. Rev. Lett. {\bf 39}, 1506 (1977).
								
\bibitem{Iwa77} S.~Iwao, Lett. Nuovo Cimento {\bf 19}, 647 (1977).

\bibitem{Gat78} R.~Gatto and F.~Paccanoni,
								Nuovo Cimento Soc. Ital.Fis {\bf 46} A, 313 (1978).								
								
\bibitem{Bre89} T.~Bressani and F.~Iazzi,
								Nuovo Cimento Soc. Ital.Fis {\bf 102} A, 597 (1989).

\bibitem{Bro90} S.~J.~Brodsky, I.~Schmidt, and G.~F.~de T\'eramond,
								Phys. Rev. Lett. {\bf 64}, 1011 (1990).

\bibitem{Bro97} S.~J.~Brodsky and G.~A.~ Miller,
								Phys. Lett. B {\bf 412}, 125 (1997).

\bibitem{Kre11} G.~Krein, A.~W.Thomas, and K.~Tsushima,
								Phys. Lett. B {\bf 697}, 136 (2011).

\bibitem{Tsu11} K.~Tsushima, D.~H.~Lu, G.~Krein, and A.~W.~Thomas,
								Phys. Rev. C {\bf 83}, 065208 (2011).

\bibitem{Yok12}	A.~Yokota, E.~Hiyama, and M.~Oka,
								Prog. Theor. Exp. Phys. {\bf 2012}, 113D01 (2012).

\bibitem{Yok14}	A.~Yokota, E.~Hiyama, and M.~Oka,
								Few-Body Syst. {\bf 55}, 761 (2014).
														
\bibitem{Kaw10} T.~Kawanai and S.~Sasaki,
								\emph{Proc. Sci}, LATTICE2010, (2010) 156 .								

\bibitem{SatoYazaki} H.~Sato and K.~Yazaki, 
                                Phys. Lett. B {\bf 98}, 153 (1981).
                                
\bibitem{Cho} S.~Cho {\it et al.} (ExHIC Collaboration), 
                                Phys. Rev. Lett. {\bf 106}, 212001 (2011).                                       
                                                   
\bibitem{Wie11} U.~Wiedner ({$\bar{\rm P}$}\!ANDA Collaboration), 
								Prog. Part. Nucl. Phys. {\bf 66}, 477 (2011).

\bibitem{Hoh11} C.~H\"ohne {\it et al.}, 
								Lect. Notes Phys. {\bf 814}, 849 (2011).

\bibitem{Tam12} H.~Tamura,
								Prog. Theor. Exp. Phys. {\bf 2012},02B012 (2012).

\bibitem{Jparc} http://j-parc.jp/index-e.html

\bibitem{Tsu08}	T.~Tsunemi, http://nuclpart.kek.jp/NP08/presentations/plenary1/pdf/np08-noumi-rcnpws2.ppt.pdf
								
\bibitem{Fel12} A.~Feliciello,
								Nucl. Phys. {\bf A881}, 78 (2012).

\bibitem{Hai07} J.~Haidenbauer, G.~Krein, U.~-G.~Meissner, and A.~Sibirtsev,
								Eur. Phys. J. A {\bf 33}, 107 (2007).

\bibitem{Hai08} J.~Haidenbauer, G.~Krein, U.~-G.~Meissner, and A.~Sibirtsev,
								Eur. Phys. J. A {\bf 37}, 55 (2008).

\bibitem{Hai09} J.~Haidenbauer and G.~Krein,
								Phys. Lett. B {\bf 687}, 314 (2010).

\bibitem{Hai10} J.~Haidenbauer, G.~Krein, U.~-G.~Meissner, and L.~Tolos,
								Eur. Phys. J. A {\bf 47}, 18 (2011).

\bibitem{Car12} T.~F.~Caram\'es and A.~Valcarce,
								Phys. Rev. D {\bf 85}, 094017 (2012).
								
\bibitem{Mat86} T.~Matsui and H.~Satz,
								Phys. Lett. B. {\bf 178}, 416 (1986).

\bibitem{Cha12} S.~Chatrchyan {\it et al.} (CMS Collaboration),
								Phys. Rev. Lett. {\bf 109}, 222301 (2012).	
																	
\bibitem{Nam61} Y.~Nambu and G.~Jona-Lasinio,
								Phys. Rev. {\bf 122}, 345 (1961).

\bibitem{Nab61} Y.~Nambu and G.~Jona-Lasinio,
								Phys. Rev. {\bf 124}, 246 (1961).
										
\bibitem{Baz11} A.~Bazavov {\it et al.},
								Phys. Rev. D {\bf 85}, 054503 (2012).

\bibitem{Aar15} G.~Aarts, :: J. Phys. Conf. Ser. 706,022004(2016).

\bibitem{Ger89} P.~Gerber and H.~Leutwyler, 
								Nucl. Phys. {\bf B321}, 387 (1989).

\bibitem{Dru91} E.~G.~Drukarev and E.~M.~Levin, 
								Prog. Part. Nucl. Phys. {\bf 27}, 77 (1991).

\bibitem{Coh92} T.~D.~Cohen, R.~J.~Furnstahl, and D.~K.~Griegel, 
								Phys. Rev. C {\bf 45}, 1881 (1992).

\bibitem{Egu76} T.~Eguchi, Phys. Rev. D {\bf 14}, 2755 (1976).

\bibitem{Vog91} U.~Vogl and W.~Weise, 
								Prog. Part. Nucl. Phys. {\bf 27}, 195 (1991).

\bibitem{Kle92} S.~P.~Klevansky, 
								Rev. Mod. Phys. {\bf 64}, 649 (1992).

\bibitem{Hat94} T.~Hatsuda and T.~Kunihiro, 
								Phys. Rep. {\bf 247}, 221 (1994).

\bibitem{Bub05} M.~Buballa, 
								Phys. Rep. {\bf 407}, 205 (2005).
								
\bibitem{Kre-Sea1} G.~Krein, P.~Tang, and A.~G.~Williams,
                                Phys.\ Lett.\ B {\bf 215}, 145 (1988).
  								
\bibitem{Kre-Sea2} G.~Krein, P.~Tang, L.~Wilets, and A.~G.~Williams,
                                Nucl.\ Phys.\  {\bf A523}, 548 (1991).
  
\bibitem{QMC} K.~Saito, K.~Tsushima, and A.~W.~Thomas,
                                Prog. Part. Nucl. Phys. {\bf 58}, 1 (2007).

\bibitem{QMC-CQM} M.~Bracco, G.~Krein, and M.~Nielsen,
                                Phys. Lett. B {\bf 432}, 258 (1998).

\bibitem{Val05} A.~Valcarce, H.~Garcilazo, F.~Fern\'andez, and P.~Gonz\'alez,
								Rep. Prog. Phys. {\bf 68}, 965 (2005).

\bibitem{Vij05} J.~Vijande, F.~Fern\'andez, and A.~Valcarce,
								J. Phys. G {\bf 31}, 481 (2005).								
  
\bibitem{Suz16} K.~Suzuki, P.~Gubler, and M.~Oka,
								Phys. Rev. C {\bf 93}, 045209 (2016).

\bibitem{Latt-Deb} C.~Allton, W.~Evans, P.~Giudice, and J.~I.~Skullerud,
                                 arXiv:1505.06616.
  
\bibitem{Screen-us} J.~Vijande, G.~Krein, and A.~Valcarce,
                                 Eur. Phys. J. A {\bf 40}, 89 (2009).


\bibitem{Gar87} H.~Garcilazo, 
								J. Phys. G {\bf 13}, L63 (1987).

\bibitem{Fon13} C.~E.~Fontoura, G.~Krein, and V.~E.~Vizcarra,
								Phys. Rev. C {\bf 87}, 025206 (2013).

\bibitem{Dot13} A.~Dot{\'e}, M.~Bayar, C.~W.~Xiao, T.~Hyodo, M.~Oka, and E.~Oset,
								Nucl.Phys. {\bf A 914}, 499 (2013).

\bibitem{Yas09} S.~Yasui and K.~Sudoh,
								Phys. Rev. D {\bf 80}, 034008 (2009).

\bibitem{Yam11} Y.~Yamaguchi, S.~Ohkoda, S.~Yasui, and A.~Hosaka,
								Phys. Rev. D {\bf 84}, 014032 (2011).

\bibitem{Kho12} A.~Khodjamirian, Ch.~Klein, Th.~Mannel, and Y.~-M.~Wang, 
								Eur. Phys. J. A {\bf 48}, 31 (2012).

\bibitem{Hai14} J.~Haidenbauer and G.~Krein, 
                                Phys. Rev. D {\bf 89}, 114003 (2014).

\bibitem{Val01} A.~Valcarce, H.~Garcilazo, R.~D.~Mota, and F.~Fern\'andez, 
								J. Phys. G {\bf 27}, L1 (2001).

\bibitem{Arn00} R.~A.~Arndt, I.~I.~Strakovsky, and R.~L.~Workman, 
								Phys. Rev. C {\bf 62}, 034005 (2000).

\bibitem{Bas09} M.~Bashkanov {\it et al.} (CELSIUS/
								WASA Collaboration),
								Phys. Rev. Lett. {\bf 102}, 052301 (2009).
								
\bibitem{Aba60} A.~Abashian, N.~E.~Booth, and K.~M.~Crowe, 
								Phys. Rev. Lett. {\bf 5}, 258 (1960).
								
\bibitem{Fries} R.~Fries, V.~Greco, and P.~Sorensen,
                                Annu. Rev. Nucl. Sci. {\bf 58}, 177 (2008).

\end{thebibliography}
\end{document}